\documentclass{ws-procs9x6}
\usepackage{subfigure}

\def\pT{\mbox{$p_T$} }
\def\v2{\mbox{$v_2$} }
\def\sqrtsNN{\mbox{$\sqrt{s_{NN}}$}}

\begin{document}

\title{High \pT hadrons in Au+Au collisions at RHIC}

\author{K.~Filimonov}

\address{Lawrence Berkeley National Laboratory \\
1 Cyclotron Road \\ 
Berkeley, CA 94720, USA\\ 
E-mail: KVFilimonov@.lbl.gov}


\maketitle

\abstracts{
High \pT hadrons produced in ultra-relativistic heavy-ion collisions 
at RHIC probe nuclear matter at extreme conditions of 
high energy density. Experimental measurements in Au+Au collisions
at \sqrtsNN=130, 200 GeV establish the existence of strong medium
effects on hadron production well into the perturbative regime.
}

One of the fundamental predictions of the 
Quantum Chromodynamics (QCD) is the existence of a deconfined 
state of quarks and gluons at the energy densities above 
1 GeV/fm$^3$~\cite{Harris:1996zx}.
This strongly interacting medium, the Quark Gluon Plasma (QGP), 
may be created in the laboratory by the collision of heavy nuclei
at high energy. The current experimental program
at the Relativistic Heavy-Ion Collider (RHIC) is aimed at detecting
the new state of matter and studying its properties. 

High \pT hadrons are produced in the initial collisions of incoming
partons with large momentum transfer. Hard scattered partons fragment
into a high energy cluster (jet) of hadrons. Partons 
propagating through a dense system may interact with the surrounding medium
radiating soft gluons at a rate proportional to the energy density
of the medium. The measurements of radiative energy loss (jet quenching)
in dense matter
may provide a direct probe of the energy density\cite{Baier:2000mf}.

The large multiplicities in nuclear collisions make full jet reconstruction
impractical. Correlations of high \pT hadrons in pseudorapidity and azimuth
allow the identification of jets on a statistical basis. 
First hints of jets at RHIC came from the two-particle azimuthal correlations
of high \pT charged hadrons in Au+Au collisions 
at \sqrtsNN=130 GeV \cite{Adler:2002ct}.
Similar analyses performed for \sqrtsNN=200 GeV \cite{Adler:2002tq,Chiu:2002ma}
directly show that hadrons at $p_T>3-4$ GeV/c result from the fragmentation
of jets.

\begin{figure}[ht]
\centering
\mbox{
\subfigure{\includegraphics[height=0.43\textwidth]{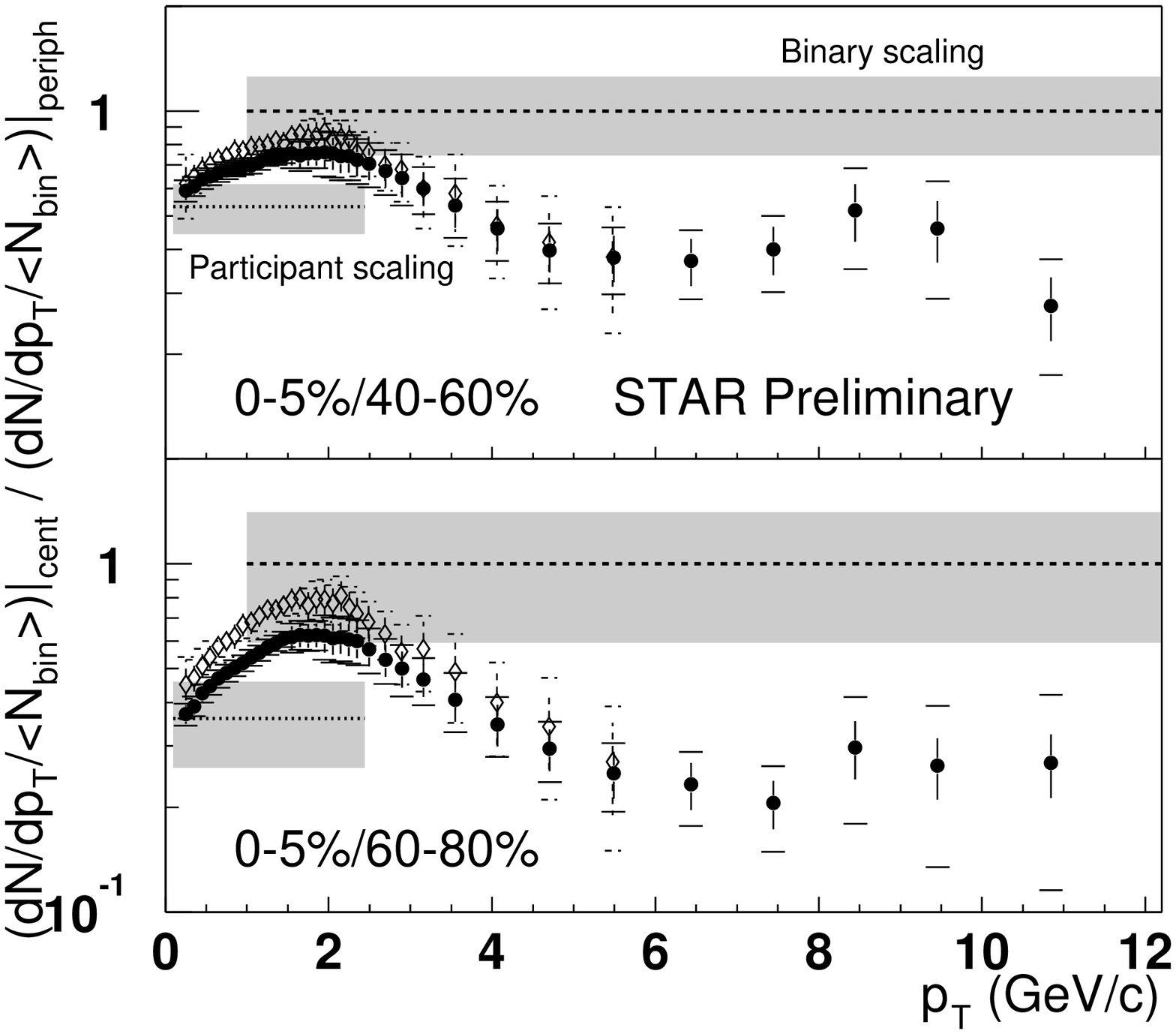}}
\subfigure{\includegraphics[height=0.43\textwidth]{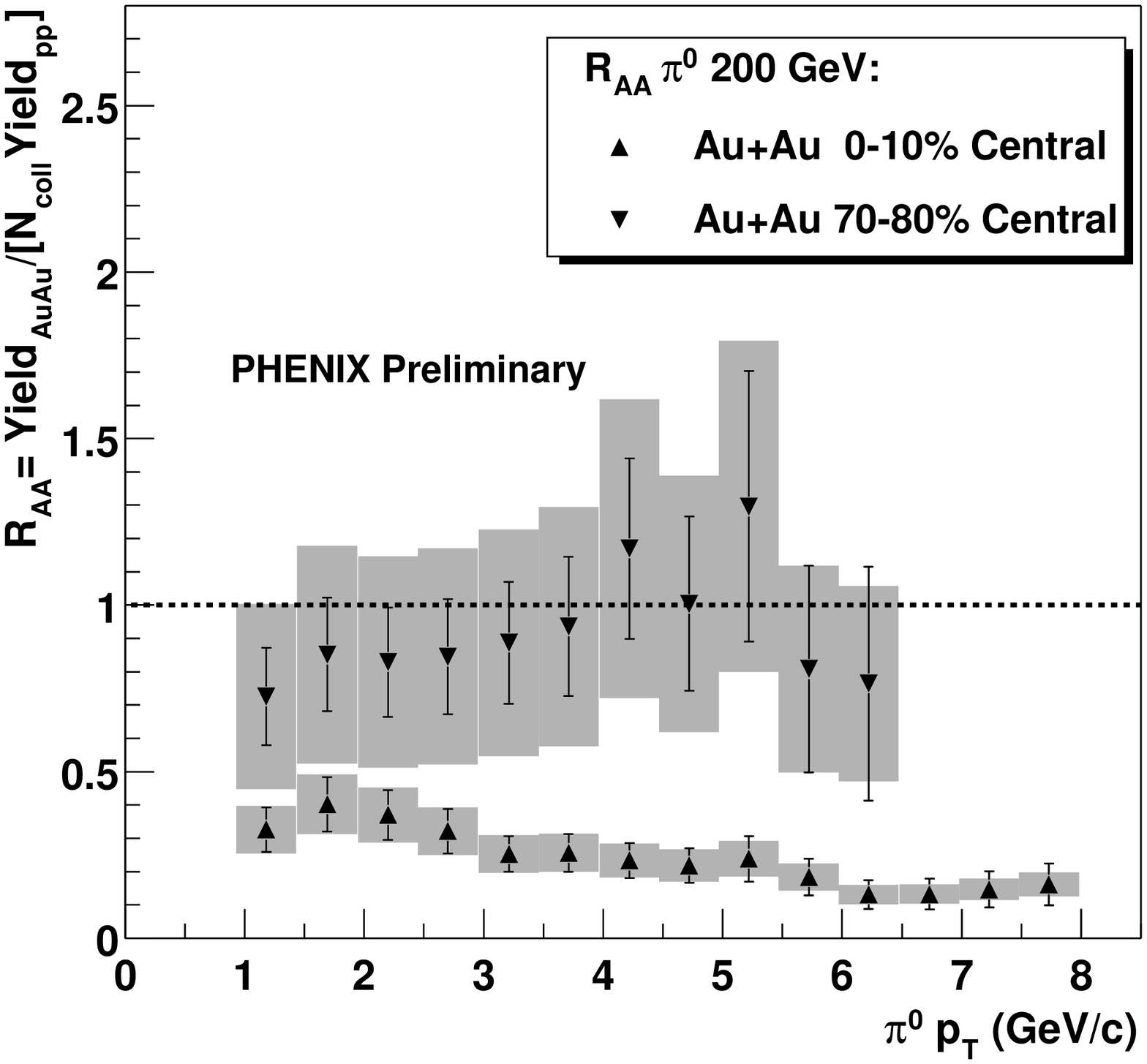}}
}
\caption{Suppression for charged hadrons (left panel, STAR) and $\pi^0$ (right panel, PHENIX). \label{RAA}}
\end{figure}

Hadrons from jet fragmentation may carry a large fraction
of jet momentum (leading hadrons). In the absence of nuclear medium
effects, the rate of hard processes should scale with the number of
binary nucleon-nucleon collisions. The yield of leading hadrons measured
in Au+Au collisions at \sqrtsNN=130 GeV 
has been shown to be significantly suppressed\cite{Adcox:2001jp,Adler:2002xw},
indicating substantial in-medium interactions. 
The high statistics 200 GeV data extended the measurements of hadron
spectra to $p_T=$12 GeV/c. Fig.\ref{RAA} shows the suppression of charged 
hadrons in central
Au+Au collisions relative to scaled peripheral Au+Au collisions 
(STAR\cite{Klay:2002xj}) and of neutral pions in central
Au+Au collisions relative to scaled p+p collisions
(PHENIX\cite{Mioduszewski:2002wt}). Suppression factors of 4-5 are observed 
for both $\pi^0$s and charged hadrons, with weak dependence on \pT at
$p_T>$5 GeV/c. The recent pQCD calculations incorporating effects of 
Cronin enhancement, nuclear shadowing and energy loss may 
describe the observed suppression\cite{Vitev:2002pf}, but a
disentangling of the effects awaits the upcoming data from d+Au collisions. 

The initial spatial almond-shaped geometry of the reaction zone in 
non-central nuclear collisions can be used to study the propagation
of partons and/or their fragmentation products through the 
azimuthally asymmetric system. 
The azimuthal anisotropy of final
state hadrons in non-central collisions is quantified by the
coefficients of the Fourier decomposition of the azimuthal particle
distributions, with the second harmonic coefficient $v_2$ referred to
as elliptic flow. The elliptic flow measurements for $p_T<2$ GeV/c 
agree in detail with hydrodynamic calculations\cite{Ackermann:2000tr}.
\begin{figure}[t]	
\centering	
\mbox{
\subfigure{\includegraphics[height=0.38\textwidth]{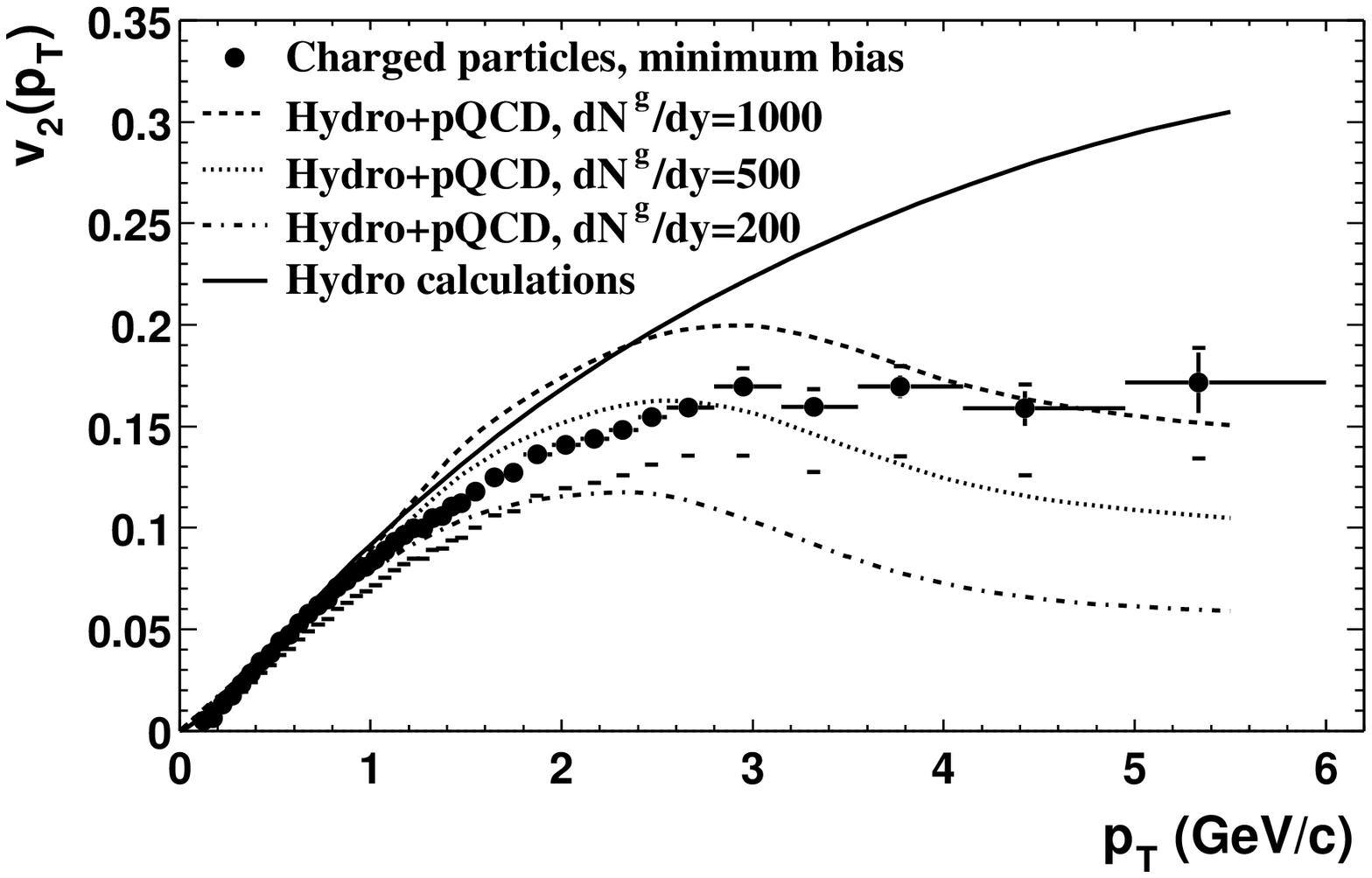}}
\subfigure{\includegraphics[height=0.38\textwidth]{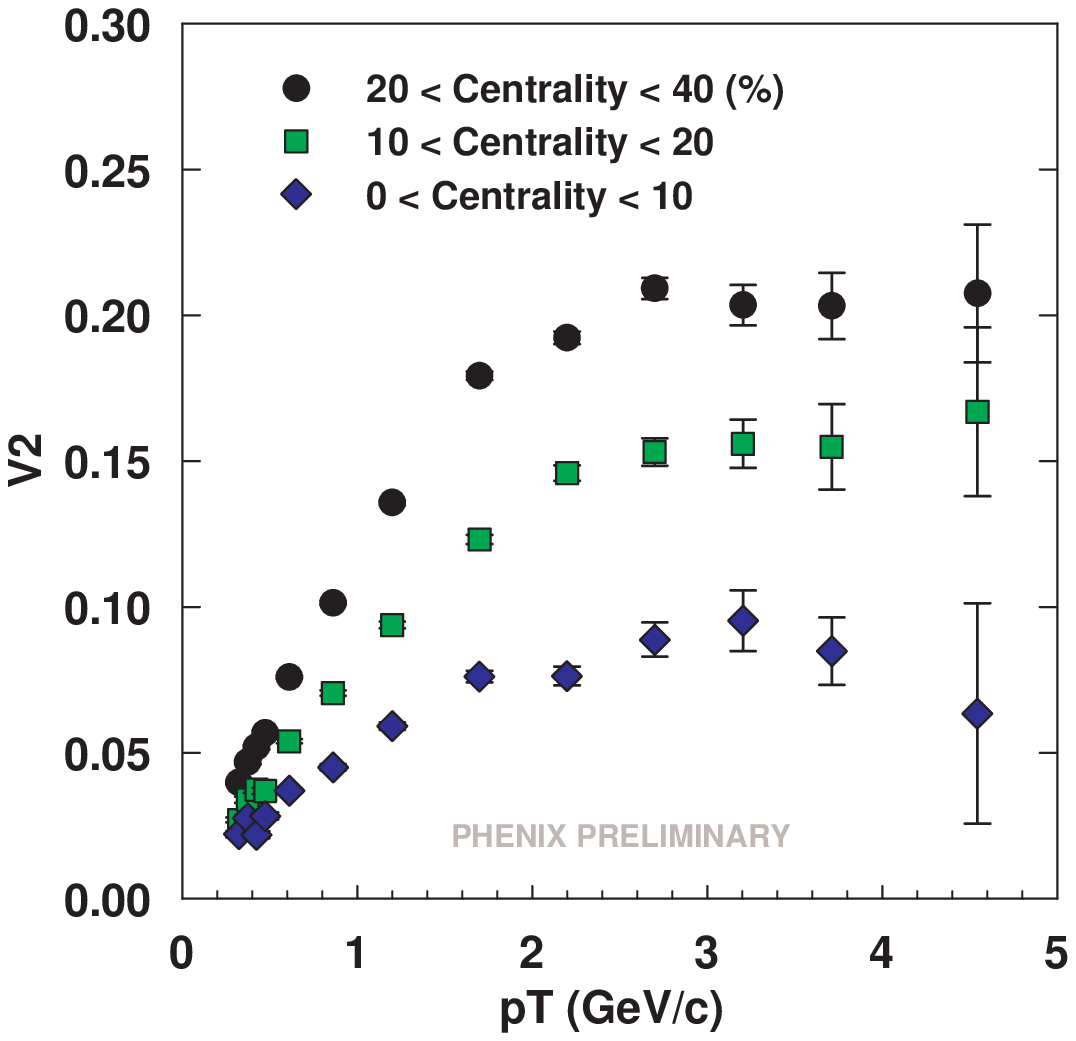}}
}
\caption{Elliptic flow $v_2(p_T)$ for charged hadrons. Left panel: minimum bias at 130 GeV (STAR) compared to hydro and pQCD calculation. Right panel: centrality dependence at 200 GeV (PHENIX).\label{v2}}
\end{figure}
Fig.\ref{v2} (left panel) shows the elliptic flow $v_2$ as a function
of $p_T$ for Au+Au collisions at 130 GeV \cite{Adler:2002ct}. Elliptic
flow rises almost linearly with transverse momentum up to 2 GeV/c,  
behavior that is 
well described by hydrodynamic calculation. Above $p_T\sim 2$ GeV/c,
$v_2(p_T)$ deviates from a linear rise and saturates for $p_T>3$ GeV/c.
The azimuthal anisotropies measured at $p_T=4-6$ GeV/c are in
qualitative agreement with the pQCD calculations including 
energy loss\cite{Gyulassy:2000gk}. Fig.\ref{v2} (right panel) 
shows the centrality
dependence of $v_2(p_T)$ measured at 200 GeV by the PHENIX Collaboration\cite{Ajitanand:2002qd}.
Finite values of $v_2$ persist to $p_T=4.5$ GeV/c for all centralities.
The results from the STAR Collaboration show the finite
values of $v_2$ up to $p_T<10$ GeV/c for non-central 
collisions\cite{Filimonov:2002xk}. However, at present
the measured values of $v_2$ contain 
a non-flow component which at high \pT comes
from intra-jet correlations.
A quantitative understanding of this effect at the highest \pT
is still needed.

\begin{figure}[ht]
\centering
\mbox{
\subfigure{\includegraphics[width=0.49\textwidth]{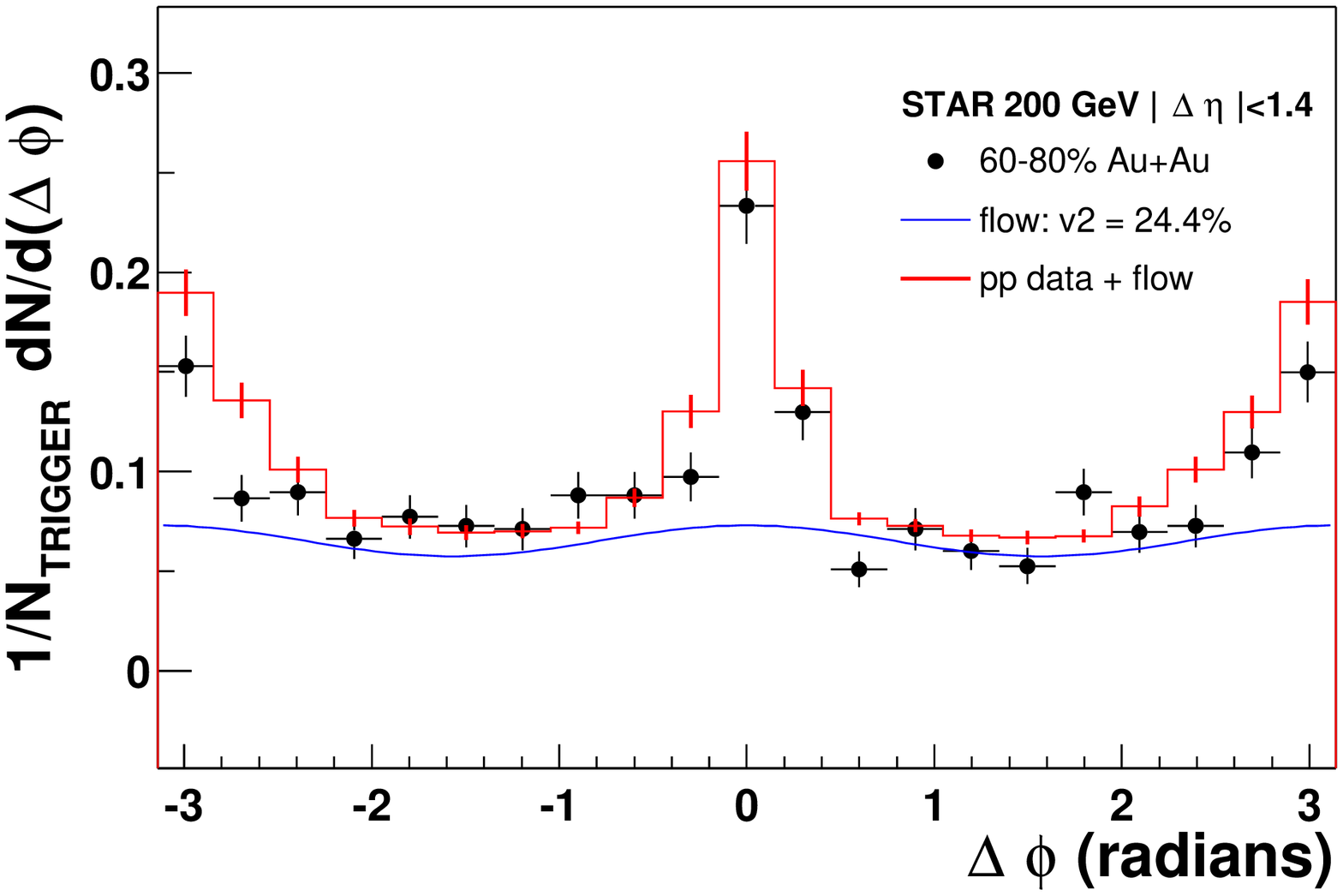}}
\subfigure{\includegraphics[width=0.49\textwidth]{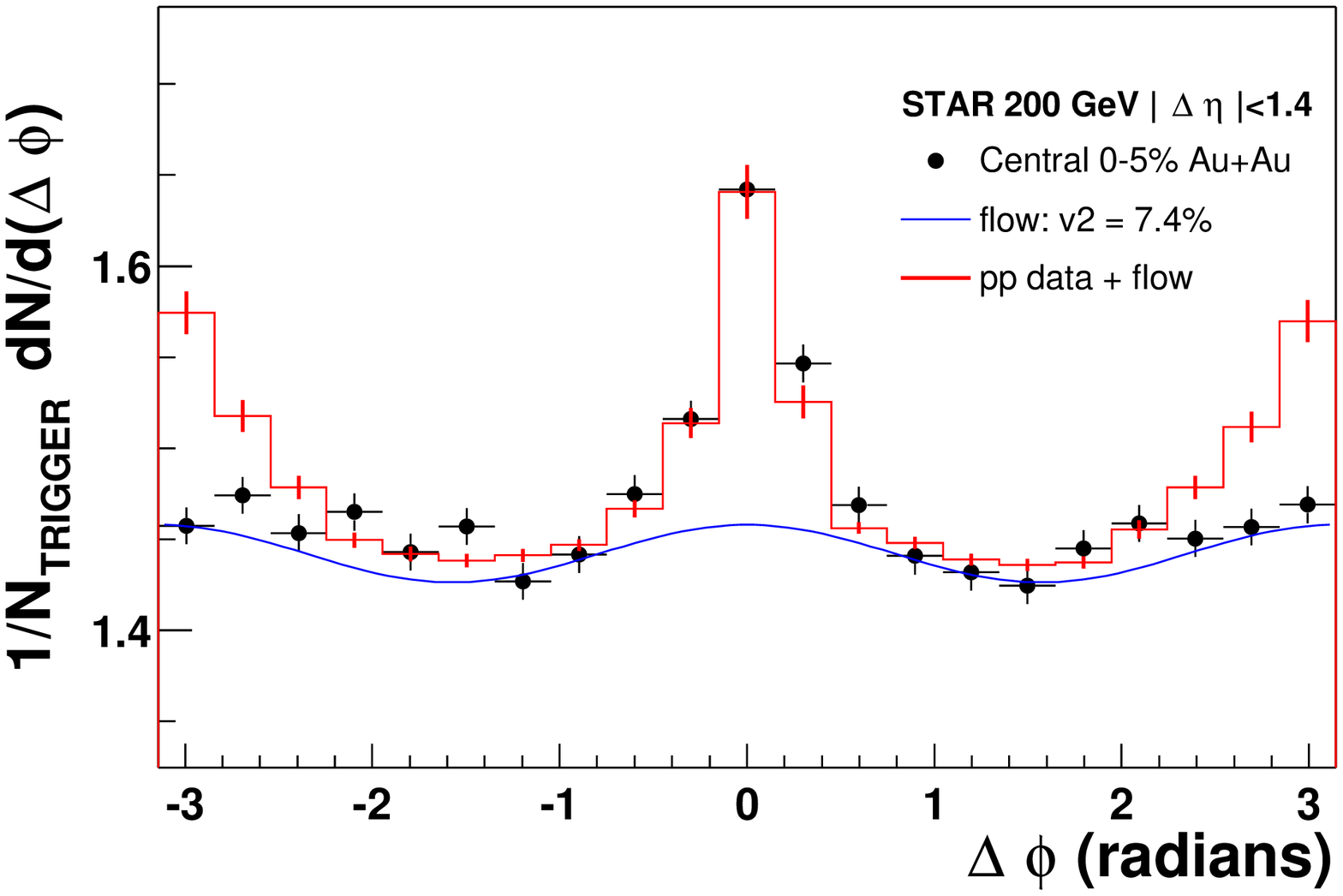}}
}
\caption{Azimuthal correlations of high \pT\ hadron pairs in Au+Au compared to p+p plus elliptic flow (STAR). Left panel: peripheral collisions. Right panel: central collisions.\label{jets}}
\end{figure}

Additional striking evidence of in-medium effects on high $p_T$ particle
production comes from the analysis of back-to-back jet correlations by 
the STAR Collaboration\cite{Adler:2002tq,Hardtke:2002ph}.
Fig.\ref{jets} shows the azimuthal angular distribution between pairs
of high \pT hadrons for the peripheral and most central Au+Au collisions
at \sqrtsNN=200 GeV. The strength of near-side correlations for both 
centralities is consistent with that measured in p+p collisions.
The away-side (back-to-back) correlations in peripheral Au+Au
collisions may be described by an incoherent superposition of jet-like
correlations measured in p+p and elliptic flow. However, back-to-back
jet production is strongly suppressed in the most central Au+Au collisions.
This indicates a substantial interaction as the hard-scattered partons
or their fragmentation products traverse the medium.

Up to now, the high \pT data from RHIC have provided a lot of 
exciting results. What is missing, however, is a coherent theoretical
description of the experimental data. Many different 
approaches are taken in attempt to describe the data, such as the pQCD with
energy loss, gluon saturation, parton cascade, surface emission, etc.
The upcoming d+Au run at RHIC will shed light on the magnitude of initial state
effects and will help to rule out certain theoretical scenarios.

\end{document}